\documentclass[preprint,showkeys,aps,prb]{revtex4}
\usepackage[dvips]{graphicx}
\begin{document}

\title{
The Equivalence of Dissipation from Gibbs' Entropy Production with
Phase-Volume Loss in Ergodic Heat-Conducting Oscillators
}

\author{
Puneet Kumar Patra, Advanced Technology Development Center         \\
Department of Civil Engineering, Indian Institute of Technology    \\
Kharagpur, West Bengal, India 721302 .                             \\
William Graham Hoover and Carol Griswold Hoover                    \\
Ruby Valley Research Institute                                     \\
Corresponding Author email : hooverwilliam@yahoo.com               \\
Ruby Valley Research Institute                                     \\
Highway Contract 60, Box 601, Ruby Valley, Nevada 89833, USA ;     \\
Julien Clinton Sprott, Department of Physics                       \\
University of Wisconsin - Madison, Wisconsin 53706, USA ;          \\
}

\date{\today}


\vspace{0.1cm}

\begin{abstract}
Gibbs' thermodynamic entropy is given by the logarithm of the phase volume,
which itself responds to heat transfer to and from thermal reservoirs.  We
compare the thermodynamic dissipation described by [ 1 ] phase-volume loss
with [ 2 ] heat-transfer entropy production.  Their equivalence is documented
for computer simulations of the response of an ergodic harmonic oscillator to
thermostated temperature gradients. In the simulations one or two thermostat
variables control the kinetic energy or the kinetic energy and its fluctuation. All of
the motion equations are time-reversible.  We consider both  {\it strong} and
{\it weak} control  variables.  {\it In every case the time-averaged
dissipative loss of phase-space volume coincides with the entropy
produced by heat transfer}.  Linear-response theory nicely reproduces the
small-gradient results obtained by computer simulation. The thermostats
considered here are {\it ergodic} and provide simple dynamical models,
some of them with as few as {\it three} ordinary differential equations,
while remaining capable of reproducing Gibbs' canonical phase-space
distribution and precisely consistent with irreversible thermodynamics.
\end{abstract}

\maketitle

\section{Historical Background and Notational Glossary}

The Irish Physicist William Rowan Hamilton formulated mechanics in terms of
a function ${\cal H}(q,p)$ depending upon the coordinates $\{ \ q \ \}$ and
momenta $\{ \ p \ \}$ describing a mechanical system.  The derivatives of this
``Hamiltonian function'' describe the system's time evolution, reducing
mechanics to the solution of $2N$ ordinary differential equations, one for the
coordinate and one for the momentum of each of the $N$ $(q,p)$ degrees of
freedom :
$$
\{ \
\dot q = +(\partial {\cal H}/\partial p) \ ; \
\dot p = -(\partial {\cal H}/\partial q) \ \} \ . 
$$
The English and German Scientists James Clerk Maxwell and Ludwig Boltzmann
showed that the thermodynamic {\it temperature} $T$ of an idealized $D$-dimensional
dilute gas composed of $N$ mass points ( mass $m$ ) is a measure of the
time averaged kinetic energy of the moving points ,
$$
\langle \ K(p)/N \ \rangle = \langle \ m\dot q^2/2 \ \rangle =
\langle \ p^2/2m \ \rangle = (D/2)kT \ [ \ {\rm ideal \ gas} \ ] \ .
$$
The proportionality constant relating the average  kinetic energy $\langle \ K
\ \rangle $ to the temperature $T$ is ``Boltzmann's Constant'' $k$ and the most
likely distribution of the momenta is given by the ``Maxwell-Boltzmann''
distribution :
$$
{\rm probability}(p) = f_{\rm MB}(p)\propto e^{ -(p^2/2mkT) } \ .
$$

The three men showed that the most likely probability distribution for any Hamiltonian
system which interacts weakly with a ``heat reservoir'' at temperature $T$ has Gibbs'
``canonical'' form. Gibbs' ``canonical distribution'' is exponential in the system
energy ( the Hamiltonian ) .  This exponential form for the probability maximizes the
thermodynamic entropy $S$ :
$$
f_{\rm canonical}(q,p)= {\rm probability}(q,p) \propto e^{-{\cal H}/kT} \ ;
\ S = k\langle \ \ln f \ \rangle \ ,
$$
where the average indicated by the angular brackets is carried out over all of the
$(q,p)$ ``states'' available to the system.  Entropy is the thermodynamic state
function associated with heat transfer.

In the present work we consider the simplest possible thermodynamic system, a
one-dimensional harmonic oscillator. For the oscillator, with force
constant and mass both  equal to unity the Hamiltonian is
${\cal H}(q,p) = (q^2/2) + (p^2/2)$
so that the motion is sinusoidal in the time $t$ , $q \propto e^{it}$, the
velocity distribution is Gaussian, $f(p) \propto e^{-p^2/2T}$, and the entropy
$S$ has a logarithmic dependence on the temperature.

Classical Hamiltonian mechanics traces out constant-energy constant-entropy orbits,
ellipses in the oscillator's $(q,p)$ space. The more general mechanics described in
the present work seeks out two
orbit types : [ 1 ] chaotic {\it equilibrium} orbits which trace out Gibbs' Gaussian distribution
as well as [ 2 ] chaotic {\it nonequilibrium} orbits where temperature depends upon
the coordinate $q$ leading to overall hot-to-cold heat transfer and to an increase
of entropy with time as is described by the Second Law of Thermodynamics,
$\dot S \ge 0 $ . Gibbs' 1902 monograph ``Elementary Principles of Statistical
Mechanics'' describes the usual classical textbook approach.  It is available free on the
internet at archive.org .

In our novel approach to nonequilibrium processes Hamilton's equations are modified by
adding time-reversible frictional forces representing the interaction of the system
with external thermostats.  In a stationary nonequilibrium process the energy
transferred {\it to} the external thermostats {\it from} the system increases the
thermostat entropy.  The corresponding entropy {\it decrease} is localized in the
system and describes the collapse of system states to a multifractal strange
attractor. In irreversible thermodynamics the reservoirs' entropy increase is
often attributed to an ``entropy production'' localized within the nonequilibrium
system and transferred to the external thermostat.  The actual decrease of system
states can be described by its Lyapunov spectrum $\{ \ \lambda \ \}$ , making contact
with the dyanmical systems research literature.  Despite the time reversibility of
the motion equations the steady-state {\it nonequilibrium} Lyapunov spectrum 
exhibits symmetry breaking.  The spectrum is {\it dissipative}, with
a negative sum, $\sum \lambda_i = -(\dot S/k)$ where $\dot S$ is the (positive)
entropy production.

\section{Introduction}

We discuss the time-reversibility and thermodynamic dissipation of several
harmonic-oscillator models, all of them {\it extensions} of the thermostated
canonical-ensemble dynamics pioneered by Shuichi Nos\'e in 1984\cite{b1,b2}.
All the resulting extended models\cite{b3,b4,b5,b6,b7,b8,b9,b10,b11,b12,b13}
studied here are chaotic and ergodic. They generate phase-space distributions
matching Gibbs' canonical distribution, Gaussian in the oscillator
coordinate $q$ and momentum $p$ with halfwidths corresponding to the
kinetic temperature $T$ .

Our {\it nonequilibrium} extensions of these equilibrium models result when
the thermostat temperature has a spatial gradient with $T = T(q)$ .  All
such nonequilibrium models discussed here generate heat flows obeying the
Second Law of Thermodynamics.  All these nonequilibrium models generate
{\it fractal} rather than smooth phase-space distributions.  The fractals'
time dependence chronicles the penetration of the fractal character to
smaller and smaller length scales with passing time, and is fully consistent
with Gibbs' phase-volume definition of entropy.

We begin with a brief discussion of time reversibility and ergodicity in
Section III.  Section IIII provides a historical sketch of time-reversible
thermostat models from Nos\'e's work to the present.  Section V
illustrates the time-reversibility of the models in nonequililbrium
stationary flows and demonstrates the consistency of all the thermostat
models with Gibbs' statistical thermodynamics.  Section VI illustrates
the consistency of these steady flows with Green and Kubo's treatment of
near-equilibrium linear-response theory.  We consider the details of the
linear-response approach for two models\cite{b7,b11}. Our Summary and
Historical Perspective Section VII includes our main conclusion from this
work: useful computational thermostats can be and should be chosen so
that the thermodynamic dissipation away from equilibrium is consistent
with the Second Law of Thermodynamics where the entropy corresponds to
Gibbs' phase-volume definition.  We relate this finding to the history
of understanding microscopic systems through the computational study of
small-system dynamics.

\section{Time-Reversible Ergodicity At and Away from Equilibrium}

Thirty years ago Nos\'e and Hoover developed two new mechanics formally
consistent with Gibbs' canonical ensemble \cite{b1,b2,b3,b4}.  These
modern mechanics share two fundamental characteristics of their
Hamiltonian ancestor, being both deterministic and {\it time-reversible}.
Any sequence of successive frames of a Nos\'e or Nos\'e-Hoover movie
played ``backward'', with the frames in reversed order, shows a reversed
motion described by exactly the same motion equations but with reversed
velocities. Hamiltonian mechanics shares this same time-reversibility
property.  

The harmonic oscillator provides the simplest example of reversibility.
If we choose a harmonic oscillator with unit mass and spring constant
any ``forward'' orbit ( with $-\tau < t < +\tau$ ) can be paired with
a time-reversed backward twin with the reversal occurring at time $t=0$ .
For instance :
$$
\{ \ q = \pm \sin(t) \ ; \ p = \pm \cos(t) \ \} \longleftrightarrow
\{ \ q = \mp \sin(t) \ ; \ p = \mp \cos(t) \ \} \ .
$$
Both orbits satisfy Hamilton's equations
$\{ \ \dot q = +p \ ; \ \dot p = -q \ \}$ .
In this simplest case the reversed version is also a mirror image of the
original, with both $q$ and $p$ changed in sign. In both cases, forward
and backward, time {\it increases}.  This corresponds to a positive timestep
$dt > 0$ in a numerical simulation.  We distinguish this physical version of
``time reversibility'' from its mathematical cousin where $dt$ changes sign
while $q$ and $p$ do not.

Nos\'e sought out a dynamics which would explore the $(q,p)$ phase space with
a probability density approaching Gibbs' canonical distribution,
$f(q,p) \propto e^{-{\cal H}(q,p)/kT}$ .
Both the Nos\'e and the simpler Nos\'e-Hoover thermostat algorithms lacked
the ergodicity required to reproduce all of Gibbs' canonical distribution for
the prototypical one-dimensional harmonic oscillator\cite{b3}. About a decade
later three more-complex algorithms, {\it doubly-thermostated} with four
motion equations rather than singly-thermostated with three, were developed.
All three have been shown to provide ergodicity for the
oscillator\cite{b5,b6,b7,b8}.  How is this ergodicity demonstrated ?

First of all ergodic motion equations necessarily satisfy the {\it stationary}
version of Liouville's continuity equation :
$$
(\partial f/\partial t) = -\nabla_r \cdot (fv) \equiv 0 \ .
$$
Abbreviate the Nos\'e-Hoover motion equations for an oscillator by a introducing
a generalized velocity $v$ for the three-dimensional flow :
$$
v = \dot r = (\dot q,\dot p,\dot \zeta) \longleftarrow \{ \ \dot q = +p \ ; \
\dot p = -q - \zeta p \ ; \ \dot \zeta = (p^2/T) - 1 \ \} \ [ \ {\rm NH} \ ] \ ,
$$
where the stationary distribution $f$ is proportional to
$e^{-q^2/2T}e^{-p^2/2T} e^{-\zeta^2/2}$ . The four nonvanishing contributions to
$(\partial f/\partial t)$ are :
$$
-\dot q(\partial f/\partial q) = p(q/T)f \ ; \
-\dot p(\partial f/\partial p) = (-q-\zeta p)(p/T)f \ ; 
$$
$$
-\dot \zeta(\partial f/\partial \zeta) = [ \ (p^2/T) - 1 \ ](\zeta)f \ ; \
-f(\partial \dot p/\partial p) = f\zeta \ .
$$
These four terms do sum to zero, showing that the motion equations are consistent
with the assumed Gaussian distribution.  The Nos\'e-Hoover equations are {\it not}
ergodic so that the vanishing of $(\partial f/\partial t)$ is not {\it sufficient}
for ergodicity.  In fact, numerical work shows that only a bit less than six
percent of the Gaussian oscillator measure is mixing and chaotic.  The remaining
94 percent is made up of regular tori, showing that the Nos\'e-Hoover distribution
is not ergodic. See {\bf Figure 1} for a cross-sectional view of the Nos\'e-Hoover
oscillator's chaotic sea.

We use the term ``chaotic'' in the usual sense here, to indicate that the
maximum Lyapunov exponent has a longtime positive average value.  Numerical
methods for measuring Lyapunov exponents so as to characterize chaos make
up a vast literature readily accessible through Wikipedia.

\begin{figure}
\includegraphics[width=4.5in,angle=-90.]{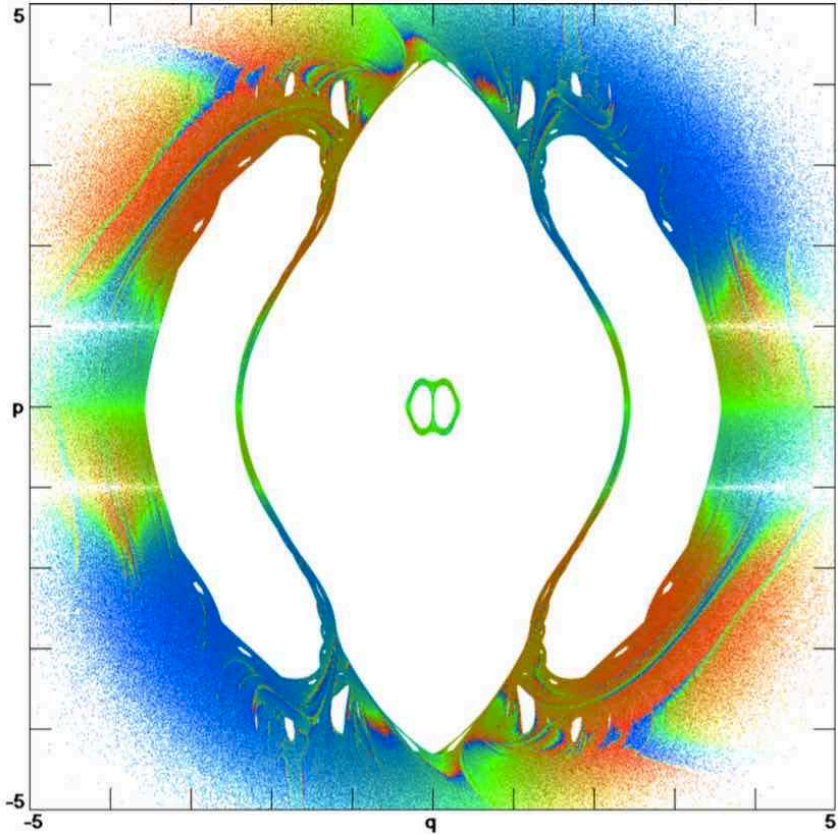}
\caption{
Penetrations of the $(q,p,0)$ plane for the chaotic Nos\'e-Hoover
oscillator with initial condition $(0,5,0)$, using points from a fourth-order
Runge-Kutta integration with a timestep $dt = 0.001$ .  Red and blue correspond
to the most positive and most negative Lyapunov exponents.  Notice the lack of
symmetry about the horizontal axis despite the time-reversibility of the equations
of motion, showing that the Lyapunov exponents' dependence on past history differs
from their relation to the unforseeable future.  This cross section of the chaotic
sea corresponds to about six percent of the Nos\'e-Hoover oscillator's Gaussian
measure.\\
}
\end{figure}

\noindent
Ergodic motion equations must necessarily reproduce the canonical moments
of the Maxwell-Boltzmann velocity distribution. With $mkT$ chosen equal to
unity to set the temperature scale, the values appropriate for Cartesian
coordinates ,
$$
\langle \ p^{2,4,6,\dots} \ \rangle _{MB} = 1, 3, 15,\dots
$$
can readily be verified from numerical simulations.  But distributions
which are ``almost'' ergodic ( for some specific \ae sthetic examples see
Figures 2-4 in Reference 13 ) can exhibit deviations so small as to be
masked by thermal fluctuations.

Two better checks of ergodicity have been implemented. Cross sections [ such
as the $(q,p,0)$ points shown in {\bf Figure 1} ], where $\zeta = 0$ or where
$\zeta = \xi = 0$ if two thermostat variables are used can be inspected
visually for the tell-tale holes indicating regular toroidal solutions within
the chaotic sea.

Additionally, the mean value of the largest Lyapunov exponent $\lambda_1$
( the longtime averaged rate of separation of two nearby trajectories,
positive for chaos and zero for tori ) can be estimated for simulations
using millions or billions of randomly chosen initial conditions.  For an
{\it ergodic} system the results cluster around a unique positive longtime
average, $\langle \ \lambda_1(t) \ \rangle \simeq \lambda_1$.  For a toroidal
system the averaged results instead cluster about zero.

The three  criteria [ moments, telltale holes, Lyapunov exponent ] have
been applied to the thermostats described in the following Section leading
to the conclusion that many different one-thermostat and two-thermostat
systems {\it are} ergodic.  Let us detail four such systems next.

\section{Deterministic Time-Reversible Thermostats (1984-2015)}

As recently as early 2015 it was thought that four or more
ordinary differential equations were required for oscillator ergodicity.
Reference 8 deals with techniques for demonstrating ergodicity as applied
to the Martyna-Klein-Tuckerman\cite{b5}, Ju-Bulgac\cite{b6}, and
Hoover-Holian\cite{b7} thermostated oscillators. For a more comprehensive
treatment see References 9 and 10.  The three thermostat types, MKT, JB,
and HH, produce chaotic dynamics $(\dot q,\dot p,\dot \zeta, \dot \xi)$
which pass visual ergodicity tests. All three of them also closely
reproduce the Cartesian velocity moments $\langle \ p^{2,4,6} \ \rangle$
characterizing the equilibrium  Maxwell-Boltzmann distribution.  Let us
begin by  reviewing the structure of these three thermostat types. 

\subsection{The Martyna-Klein-Tuckerman ``Chain'' Thermostat (1992) }
The Martyna-Klein-Tuckerman thermostat uses two control variables, $\zeta$
and $\xi$, with $\zeta$ controlling $\langle \ p^2 \ \rangle$ and $\xi$ 
controlling $\langle \ \zeta^2 \ \rangle$ :
$$
\{ \ \dot q = p \ ; \ \dot p = -q - \zeta p  \ ; \ \dot \zeta = (p^2/T)
- 1 - \xi \zeta \ ; \ \dot \xi = \zeta^2 - 1 \ \} \ [ \ {\rm MKT} \ ] \ .
$$
The steady-state distribution corresponding to these oscillator motion
equations is an extension of Gibbs' canonical one :
$$
f_{\rm MKT}(q,p,\zeta,\xi) \propto
e^{-q^2/2T}e^{-p^2/2T}e^{-\zeta^2/2}e^{-\xi^2/2} \longrightarrow \
$$
$$
(\partial f/\partial t) = -\nabla_r\cdot (fv) \equiv 0 \ {\rm where}
\ v = \dot r \equiv (\dot q,\dot p,\dot \zeta,\dot \xi) \ .
$$
The {\it stationarity} test from the continuity equation,
$(\partial f/\partial t) = 0$ , provides a necessary, but not necessarily
sufficient, condition that {\it any} set of motion equations must satisfy for
ergodicity. Martyna, Klein, and Tuckerman\cite{b5} emphasized that any number of
additional control variables can be added to form a ``chain'' of thermostats.

\subsection{The Ju-Bulgac Cubic Thermostat (1993) }
The Ju-Bulgac thermostat\cite{b6} likewise uses two control variables but
includes {\it cubic} dependences following the observation of Bauer, Bulgac,
and Kusnezov that cubic terms enhance chaos and ergodicity\cite{b6,b9,b10} :
$$
\{ \ \dot q = p \ ; \ \dot p = -q - \zeta^3 p - \xi(p^3/T) \ ;
\ \dot \zeta = (p^2/T) - 1 \ ;
\ \dot \xi = (p^4/T^2) - 3(p^2/T) \ \} \ [ \ {\rm JB} \ ] \ .
$$
The steady-state distribution here is Gaussian in $\zeta ^2$ rather than
$\zeta$ :
$$
f_{\rm JB}(q,p,\zeta,\xi) \propto
e^{-q^2/2T}e^{-p^2/2T}e^{-\zeta^4/4}e^{-\xi^2/2} \longrightarrow
(\partial f/\partial t)  \equiv 0 .
$$
At unit temperature $T=1$ the rms rate $ | \ \dot p \ |$ at which the Ju-Bulgac
momentum moves through phase space is about three times faster than that of the
simpler Martyna-Klein-Tuckerman momentum :
$$
\sqrt{\langle \ q^2 + p^2\zeta^6 + p^6\xi^2 \ \rangle } \simeq \sqrt{18.028} \
{\it versus} \ \sqrt{\langle \ q^2 + p^2\zeta^2 \ \rangle } = \sqrt{2} \ .
$$
From the numerical standpoint cubic thermostat variables enhance chaos
and mixing without incurring the considerable stiffness associated with
quintic controls.

\subsection{The Hoover-Holian Thermostat (1996)}
Like the two preceding, the Hoover-Holian thermostat\cite{b7} uses two control
variables.  The first  one is allocated to fixing the oscillator temperature
$\zeta \rightarrow \langle \  p^2 \ \rangle$ while the second fixes the
fluctuation of the temperature $\xi \rightarrow \langle \  p^4 \ \rangle$ :
$$
\{ \ \dot q = p \ ; \ \dot p = -q - \zeta p - \xi (p^3/T) \ ; \
\dot \zeta = (p^2/T) - 1 \ ; \ \dot \xi = (p^4/T^2) - 3(p^2/T) \ 
[ \ {\rm HH} \ ] \ .
$$
At unit temperature the rms rate at which the Hoover-Holian momentum moves,
$\sqrt{\langle \ q^2 + p^2\zeta^2 + \xi^2 p^6 \ \rangle }
=\sqrt{17}$ , is nearly the same as the Ju-Bulgac speed. The Hoover-Holian
thermostat variables $\zeta$ and $\xi$ exert
what we term ``strong'' control of the temperature and its fluctuation, in
that longtime averages of the thermostat motion equations constrain moments
proportional to the kinetic energy and its fluctuation :
$$
\langle \ \dot \zeta \ \rangle = 0 \longrightarrow
\langle \ (p^2/T) \ \rangle \equiv 1 \ ;
\langle \ \dot \xi \ \rangle = 0 \longrightarrow
\langle \ (p^4/T^2) \ \rangle \equiv \langle \  3(p^2/T) \ \rangle \ .
$$
These strong constraints can be applied equally well in nonequilibrium
situations.  Nonequilibrium applications of the MKT thermostat
typically lead to nonzero correlated values of the thermostat variables,
$\langle \ \zeta \xi \ \rangle$ so that the definition of the kinetic
temperature $\langle \ (p^2/T) \ \rangle \equiv 1$ is violated.

At equilibrium the steady-state distribution corresponding to the HH
motion equations is exactly the same as the Martyna-Klein-Tuckerman
four-dimensional Gaussian :
$$
f_{\rm HH}(q,p,\zeta,\xi) = f_{\rm MKT}(q,p,\zeta,\xi) \propto 
e^{-q^2/2T}e^{-p^2/2T}e^{-\zeta^2/2}e^{-\xi^2/2} \longrightarrow
(\partial f/\partial t)  \equiv 0 \ .
$$

\subsection{The Ergodic Single-Thermostat 0532 Model (2015)}

Very recently\cite{b11,b12,b13} a variety, both novel and wide, of
{\it singly}-thermostated ergodic algorithms has been developed and
applied to the one-dimensional harmonic oscillator.  The simplest of
them, the ``0532 Model'', consists of only three ordinary differential
equations for the oscillator coordinate $q$ , velocity $p$ , and
friction coefficient $\zeta$ at a thermostat temperature $T$ :
$$
\dot q = p \ ; \ \dot p = - q - \zeta[ \ 0.05p + 0.32(p^3/T) \ ] \ ; 
$$
$$
\dot \zeta = 0.05[ \ (p^2/T) - 1 \ ] + 0.32[ \ (p^4/T^2) - 3(p^2/T) \ ] \
; \ [ \ 0532 \ {\rm Model} \ ] \ .
$$
We term this simultaneous control of the second and fourth moments,
$\langle \ p^{2 \ {\rm and} \ 4} \ \rangle $, ``weak'' because a linear
combination of the moments is controlled rather than enforcing the separate
control of {\it both} moments, as in the earlier work in References 5-9 .
Numerical solutions of the 0532 oscillator model indicate that it {\it is}
ergodic and corresponds to Gibbs' canonical ensemble multiplied by a
Gaussian distribution for the thermostat control variable $\zeta $ :
$$
f_{\rm 0532}(q,p,\zeta,\xi) =  \propto
e^{-q^2/2T}e^{-p^2/2T}e^{-\zeta^2/2}  \longrightarrow                                                                    
(\partial f/\partial t)  \equiv 0 \ .
$$
Because the 0532 model motion occurs in just three dimensions rather
than four, it is well-suited to analysis.  This model, like its three
predecessors in this Section, is time-reversible, even in the nonequilibrium
case where the temperature varies in space, $T=T(q)$ .  Let us review the
reversibility property in that specific {\it nonequilibrium} case.
 
\section{Time Reversibility Away From Equilibrium -- 0532 Model}

{\it At} equilibrium the forward and backward trajectories for canonical
oscillators, using any of the four ergodic sets of motion equations,
are qualitatively much the same.  No holes in the cross sections, good values
for the even velocity moments, longtime averaged Lyapunov exponent the
same for any initial condition.  In short -- deterministic, time-reversible,
ergodic.

{\it Away from} equilibrium, thermodynamic dissipation can be investigated,
still time-reversibly, by adding a localized temperature gradient
$(dT/dq) = [ \ \epsilon/\cosh^2(q) \ ]$ enabling heat transfer through a nonzero average
current $(p^3/2)$ :
$$
1 - \epsilon < T < 1 + \epsilon = T(q) = 1 + \epsilon \tanh(q)
\longrightarrow \langle \ (p^3/2) \ \rangle < 0
\longrightarrow (\dot S/k) < 0 \ .
$$
Here $\epsilon$ is the maximum value of the temperature gradient, $T^\prime(0)$ .
The negative entropy change, causing the phase volume to shrink onto a strange
attractor is due to the net heat loss {\it from} the oscillator {\it to} the 
coordinate-dependent 0532 thermostat temperature $T(q)$ . From the standpoint of
steady-state irreversible thermodynamics the overall heat loss is offset by an
internal ``entropy production'' so that the {\it net} change of oscillator
``entropy'' vanishes.  We remind the reader that Gibbs' entropy is minus infinity
for fractal attractors so that the irreversible-thermodynamics concept of
nonequilibrium entropy is problematic.  The artificial entropy change could be also
be viewed as the result of ongoing coarse-graining ( which would artificilly
increase Gibbs' entropy ) at the level of the computational roundoff error
( in the sixteenth or seventeenth digit ).

The temperature gradient destroys the ``global [ overall ] reversibility'' of the
motion equations. Although in principle reversible the chaotic instability of the
dynamics, evidenced by a positive Lyapunov exponent, makes this ``irreversibility''
possible.  This irreversibility is evidenced by a Lyapunov spectrum with a {\it
negative} sum so that the longtime averaged distribution is a fractal strange
attractor with a reduced information dimension rather than a smooth three-dimensional
Gibbsian distribution.

Among the thermostats we have considered only the Nos\'e-Hoover equations show
that a fractal attractor is {\it not} inevitable.  In the Nos\'e-Hoover case a majority
of initial conditions give rise to phase-space tori, orbits with no longtime
tendency toward dissipation.  All of the ergodic thermostats invariably produce
small-gradient dissipation rather than tori so that their orbits exhibit what we call
``global irreversibility''.

The equilibrium ( $\epsilon = 0$ and unit temperature $T = 1$ ) Lyapunov spectrum
for the 0532 model, $\{ \ \lambda \ \} = \{ \ +0.144,0,-0.144 \ \}$ 
sums to zero corresponding to the three-dimensional Gaussian distribution,
$f \propto e^{-q^2/2}e^{-p^2/2}e^{-\zeta^2/2}$ .  The time-averaged growth
rates of infinitesimal one-, two-, and three-dimensional phase space
volumes are given by
$$
\{ \ \lambda_1, \ \lambda_1+\lambda_2, \ \lambda_1+\lambda_2+\lambda_3 \ \} \ .
$$
In the nonequilibrium
case with $\epsilon = 0.50$ the time-averaged spectrum becomes asymmetric,
$\{ \ +0.1135,0,-0.1454 \ \}$ , corresponding to the time-averaged growth of
a length or an area in phase space $\simeq e^{+0.1135t}$ but to {\it shrinkage}
of an infinitesimal three-dimensional phase volume $\otimes$ :
$$
(\dot \otimes/\otimes) = 0.1135-0.1454 = - 0.0319 \longrightarrow
D_{KY} = 2 + (0.1135/0.1454) = 2.78 \ . 
$$

\begin{figure}
\includegraphics[width=4.5in,angle=-90.]{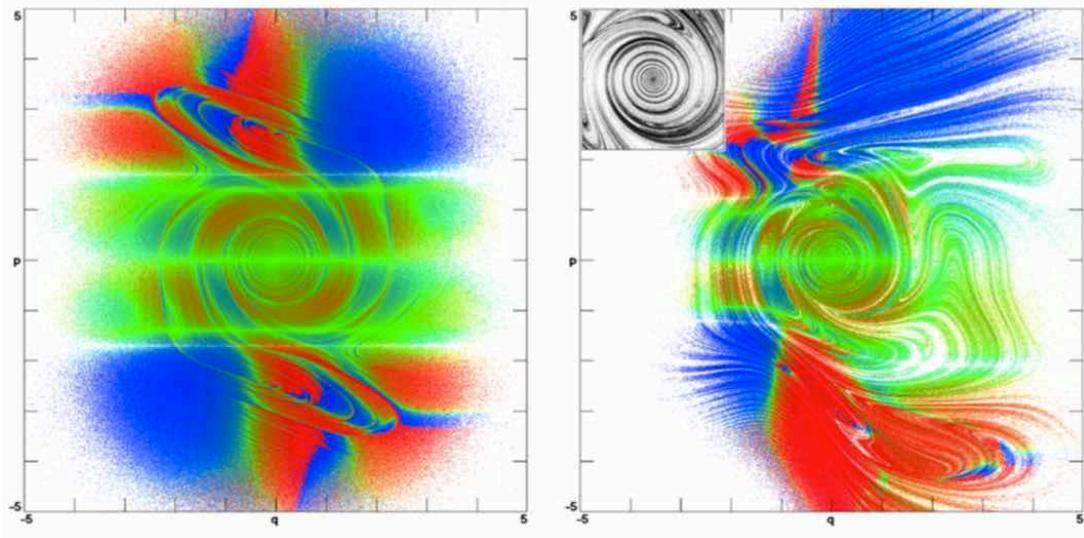}
\caption{
Penetrations of the $(q,p,0)$ plane for the chaotic and ergodic
0532 Model using fourth-order Runge-Kutta integration with a timestep
$dt = 0.001$ .  The red and blue points correspond to maximum and minimum
values of the local Lyapunov exponent.  The equilibrium $\zeta = 0$ cross
section at the left shows inversion symmetry, corresponding to viewing
the oscillator in a mirror.  The lack of symmetry about the horizonal
$p=0$ axis shows that the exponents depend upon the past rather than the
future.  The nonequilibrium section ( $\epsilon = 0.50$ ) shown to the right
displays no symmetry and is multifractal.  The black-and-white inset
shows the cross-sectional density in the $2 \times 2$ central region of the
phase-plane section.\\
}
\end{figure}

\noindent
Kaplan and  Yorke's linear interpolation predicts a strange-attractor dimension
of 2.78.  Cross sections of the equilibrium and nonequilibrium 0532 dynamics
are shown in {\bf Figure 2} . Just as at equilibrium the nonequilibrium
strange-attractor's motion equations are time-reversible.  Any forward-in-time
sequence
$\{ \ +q,+p,+\zeta \ \}$ 
corresponds to a twin sequence $\{ \ +q,-p,-\zeta \ \}$ with the order of the
$(q,p,\zeta)$ points reversed.  {\it Locally} this reversed sequence satisfies the
same equations of motion with errors of order $(dt^5/120)$ for fourth-order
Runge-Kutta integration.  But any attempt to {\it generate} such a reversed
sequence numerically fails because the Lyapunov spectrum of the reversed
sequence would correspond to $\{ \ +0.1454,0,-0.1135 \ \}$ .  The positive
exponent sum indicates an unstable {\it repellor} with a {\it diverging} phase
volume, $(\dot \otimes/\otimes) = +0.0319$ . Any attempt to follow the repellor
numerically will instead seek out the nearby attractor ( both are still ergodic,
at least if $\epsilon$ is small ) which, though unstable for a line or an area,
is less so than the repellor.  The repellor properties {\it can} ( only ) be
observed by the expedient of {\it storing} and reversing a trajectory.  The
cross section associated with a stored ten-billion-point attractor trajectory is
illustrated in {\bf Figure 3} .  Note the lack of $\pm p$ symmetry in the
coloring of the local Lyapunov exponent, $\lambda_1(t)$ .

\begin{figure}
\includegraphics[width=4.5in,angle=-90.]{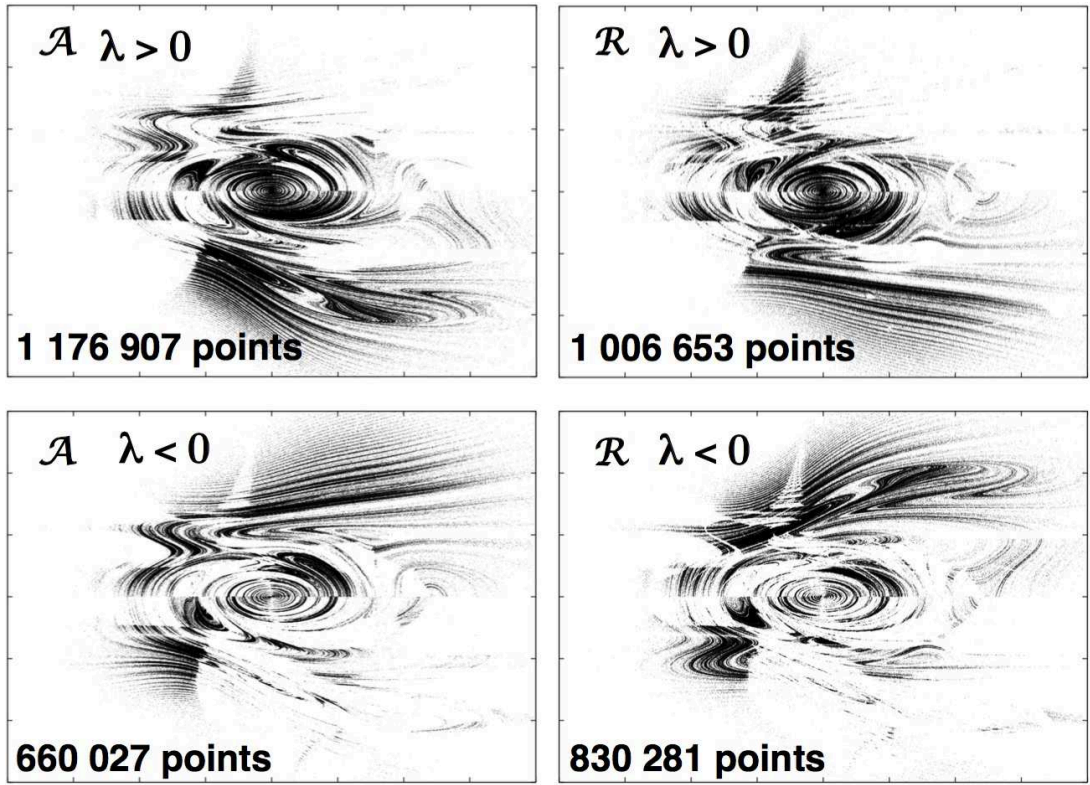}
\caption{
Penetrations of the $(q,p,0)$ plane for the chaotic and ergodic
0532 Model using a ten-billion-point attractor reference trajectory ( denoted A ) and
fourth-order Runge-Kutta integration with a timestep $dt = 0.001$ for the
satellite trajectory.  This trajectory crosses the $\zeta = 0$ plane 1 836 934 times.
The signs of the largest Lyapunov exponent at each crossing are indicated
for both the attractor and the repellor ( denoted R ).  By plotting the positive
and negative points separately the lack of any symmetry is clear. The repellor points
are identical to those of the attractor by are traced out in the opposite direction.
For both the attractor and the repellor the separatiom of the reference and satellite
trajectories is $\sqrt{(q_s-q_r)^2 + (p_s-p_r)^2 + (\zeta_s-\zeta_r)^2} = 0.000001$ .
Here the maximum value of the nonequilibrium temperature gradient is $\epsilon = 0.5$
 \ .\\
}
\end{figure}

\noindent
This instructive problem illustrates two general principles : [ 1 ] the phase
volume of the steady-state attractor is zero and singular everywhere {\it
despite the time-reversibility of the motion equations} ; [ 2 ] any typical
three-dimensional phase volume first expands and leaves the vicinity of the
( ergodic ) fractal repellor and then shrinks in order to join its
mirror-image ( and likewise ergodic ) fractal attractor exponentially fast.
Both these features correspond to the paucity of nonequilibrium states and to
the irreversibility described by the Second Law of Thermodynamics.

There is more.  Consider two additional equally-significant observations.  First,
the comoving shrinkage rate in phase space corresponds precisely and instantaneously
to the loss of Gibbs' entropy for the system.  To illustrate consider the 0532
model ,
$$
\dot q = p \ ; \ \dot p = - q - \zeta[ \ 0.05p + 0.32(p^3/T) \ ] \ [ \ 0532 \ ] \ .
$$
$$
(\dot S/k) = (\dot \otimes/\otimes) \equiv (\partial \dot q/\partial q) +
(\partial \dot p/\partial p) + (\partial \dot \zeta/\partial \zeta) =
0 - \zeta[ \ 0.05 + 0.96(p^2/T) \ ] + 0 \ .
$$
Second, this loss rate also corresponds precisely, {\it when time-averaged},
to the kinetic energy ( or heat $Q$) extracted by the thermostat forces, divided
by the thermostat temperature $T$ :
$$
\langle \ (\dot Q/T)  \ \rangle = -\langle \ \zeta[ \ 0.05(p^2/T)+0.32(p^4/T^2) \ ] \ \rangle
= \langle \ (\dot S/k) \ \rangle \ .
$$
The time-averaged value $\langle \ \zeta[ \ 0.05 + 0.96(p^2/T) \ ] \ \rangle $ ,
follows from the
time-averaged evolution equation for the squared thermostat variable $(\zeta^2/2)$ :
$$
\langle \ \zeta \dot \zeta = 0 = 0.05\zeta [ \ (p^2/T) - 1 \ ] +
0.32\zeta [ \ (p^4/T^2) - 3(p^2/T) \ ] \ \rangle \ .
$$

The time-averaged phase-volume loss, equivalent to the dissipation seen in the
heat $Q$ lost to thermal reservoirs divided by the reservoir temperature $T$ ,
$$
\langle \ (\dot Q/T) \ \rangle = \langle \ k(\dot \otimes/\otimes) \ \rangle
= \langle \ \dot S \ \rangle \ ,
$$
holds generally for {\it all} the thermostat models discussed here.  This
identity holds even for the Nos\'e-Hoover model, which is not ergodic.  It holds for
other power laws. Suppose for instance that the thermostat force is proportional
to odd powers of $\zeta$ and $p$ :
$$                                                                                                 
-A_{mn}\zeta ^{2m+1}(p^{2n+1}/T^n) \                                                                
$$
so that the equilibrium distribution is proportional to 
$$
f \propto e^{-p^2/2T}e^{-\zeta^{2m+2}/(2m+2)} \ .
$$
Gibbs' phase-space  dissipation, from $-(\partial \dot p/\partial p)$ gives
a contribution to the system entropy :
$$
(\dot S/k) = -(2n+1)A_{mn}\zeta ^{2m+1}(p^{2n}/T^n) 
$$
The entropy change from the contribution of the same dissipative term to heat
transfer is :
$$
(\dot Q/T) = -A_{mn}\zeta ^{2m+1}(p^{2n+2}/T^{n+1}) \ .
$$
A look at the equation of motion for the friction coefficient, multiplied by
$\zeta^{2m+1}$ and time averaged shows that $(\dot S/k)$ and $(\dot Q/T)$
are equivalent :
$$
\langle \ \zeta^{2m+1}\dot \zeta  \ \rangle = \langle \ \zeta^{2m+1}A_{mn}
[ \ (p^{2n+2}/T^{n+1}) - (2n+1) (p^{2n}/T^{n}) \ ] \ \rangle = 0 ,
$$
This is a consequence of the vanishing of the longtime averaged value of a
bounded quantity, in this case $(d/dt)[ \ \zeta^{2m+2}/(2m+2) \ ] \ .$
Generalized models, like the 0532 model, can use two or more power-law
contributions to thermostat forces.
This equivalence of Gibbs' entropy production with that from irreversible
thermodynamics points the  way forward toward consistent theories of
nonequilibrium steady states either near to or far from equilibrium.

In the past it has been pointed out that it {\it is} possible to develop
thermostats for which the phase-volume and heat-transfer rates are {\it not}
closely related\cite{b15,b16,b17}.  This potential loss of a family
relationship recalls Tolstoy's thought: ``All happy families are alike ;
each unhappy family is unhappy in its own way.''  We emphasize here that
the close relationship linking phase volume to thermodynamics is to be
celebrated rather than avoided.

We note that our dimensionless friction coefficients {\it could} be multiplied
by relaxation times or by powers of the temperature, changing their units.  We
have carefully chosen the forms used here in order to guarantee the consistency
of the motion equations with both Gibbs' canonical distribution and with
thermodynamics.  Dimensionless friction coefficients seem to us the simplest
approach to thermodynamic consistency. 

In the 1950s Green and Kubo showed that their ``linear-response'' theory expresses
nonequilibrium transport coefficients in terms of equilibrium correlation functions.
This same theory can be applied to the various thermostats we have described.  Next
we illustrate this idea for two examples, the doubly-thermostated Hoover-Holian
thermostat and the singly-thermostated 0532 Model.

\section{Linear Response Theory with a Temperature Gradient}

We have celebrated the equivalence of two measures of dissipation, phase volume loss
and Gibbs' entropy production when any one of our five of thermostat models
[ NH, MKT, JB, HH, 0532 ] is time averaged.  This equivalence guarantees their
usefulness in simulations consistent with dynamical equivalents of the canonical
ensemble.  Green-Kubo linear-response theory is a perturbation theory based on
Gibbs' ensembles.  Typically the energy is modified by a perturbation, giving
rise to a nonequilibrium flux.  In our case both the energy and the temperature are
modified by introducing a temperature profile along with a stabilizing frictional
force.  Let us demonstrate their theory's usefulness for the Hoover-Holian
$(q,p,\zeta\xi)$ and the 0532 Model $(q,p,\zeta)$ oscillators as two concrete examples.

\subsection{Hoover-Holian Oscillator with Temperature Gradient}

We begin with the extended canonical distribution for the oscillator with energy $E$
and at a temperature $T$ of unity :
$$
f(qp\zeta\xi)_{\rm HH} \propto
e^{-{\cal H}(q,p)/kT}e^{-\zeta^2/2}e^{-\xi^2/2} =
e^{-q^2/2T}e^{-p^2/2T}e^{-\zeta^2/2}e^{-\xi^2/2} \ .
$$
Adding a temperature perturbation ,
$$
T = 1 \longrightarrow T = 1 + \Delta T = 1 + \epsilon \tanh (q) \ ,
$$
we wish to compute the responding current, $(p^3/2)$ as a function of time.

The simplest form of the Hoover-Holian motion equations is :
$$
\{ \ \dot q = p \ ; \ \dot p = -q - \zeta p - \xi (p^3/T)  \ ; \ \dot \zeta = (p^2/T)                                                                                                           
- 1 \ ; \ \dot \xi = (p^4/T^2) - 3(p^2/T) \ \} \ [ \ {\rm HH} \ ] \ .                             
$$
The time-dependent change of the canonical weight $e^{-\Delta(E/kT)}$
can be linearized in the thermal perturbation $\epsilon$ with the result :
$$                                                                                                                                                                                      
(f_{neq}/f_{eq}) =
1 + {\int_0^t[ \ \epsilon\tanh(q) \ ]_0[ \ -\zeta p^2 -\xi (p^4/T) \ ]_{t^\prime}dt^\prime} \ .
$$
We can use this {\it nonequilibrium} perturbation to compute the current $(p^3/2)$ at
time $t$ from the {\it equilibrium} correlation function ( which depends only on the
time difference $t^\prime$ ) :
$$                                                                                                                                                                                      
\textstyle{                                                                                                                                                                             
\langle \ (p^3/2) \ \rangle_{\rm neq} = \int_0^t \langle \
[ \ \epsilon \tanh(q) \ ]_0
[ \ -\zeta p^2 -\xi (p^4/T) \ ]_0(p^3/2)_{t^\prime} \ \rangle_{\rm eq} dt^\prime                                                                                             
\ .
}                                                                                                                                                                                       
$$
 
A highly-accurate equilibrium calculation can be based on the fact that the four-dimensional
equilibrium measure is ergodic, a Gaussian probability density known in advance. To compute
averages we begin with a grid of [ $100 \times 100 \times 100 \times  100$ ]
equiprobable points and use these as the initial conditions for computing both the
nonequilibrium current and the equilibrium correlation function.  The excellent agreement
shown in {\bf Figure 4} confirms the analysis showing that both the equilibrium distribution
function and its linear perturbation are well suited to numerical exploration.  The figure
compares the linear-response expression for the current to that actually measured with
nonequilibrium molecular dynamics at the relatively small field strength $\epsilon = 0.10$ .
We conclude that simple linear-response theory is a fringe benefit of our deterministic
ergodic thermostat models.\\

\begin{figure}
\includegraphics[width=4.5in,angle=90.]{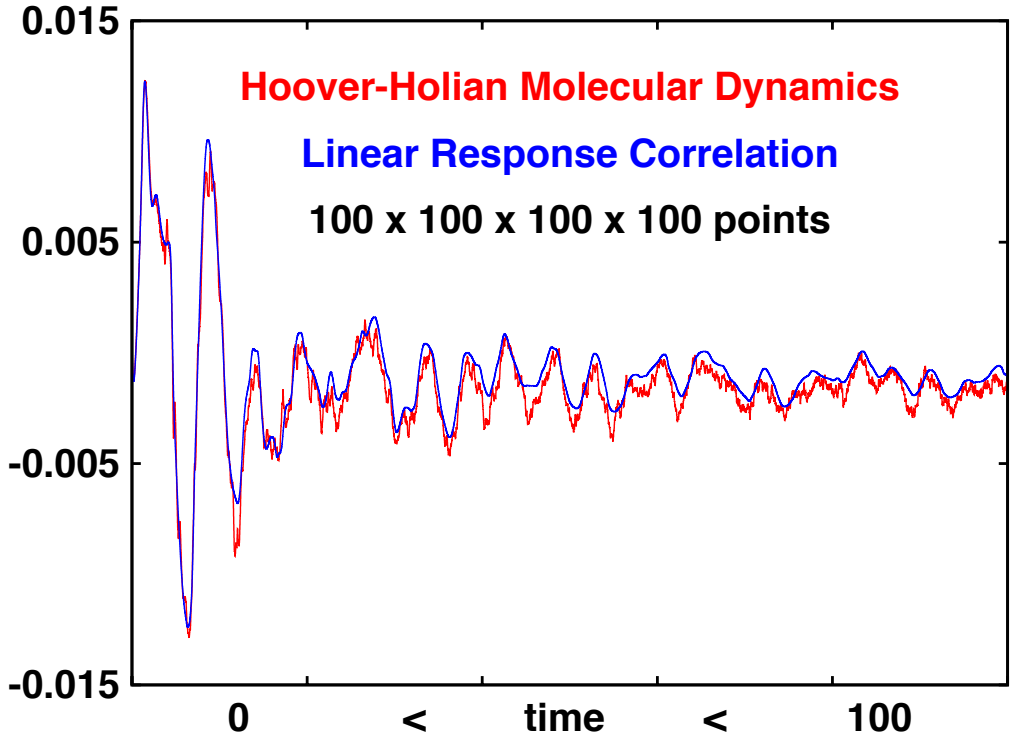}
\caption{
Comparison of the linear-response correlation function ( blue ) with the measured current
( red ) for the HH oscillator at a field strength $\epsilon = 0.10$ .  Results for
$T = 1 + 0.10 \tanh(q)$ ( shown here) and $T = [ \ 1 - 0.10 \tanh(q) \ ]^{-1}$ are very
similar and confirm that $\epsilon = 0.10$ is close to the linear regime. The phase-space
integration uses $100^4$ equally-probable Gaussian points as the initial states for the
averaged current $\langle \ (p^3/2) \ \rangle $ and for the linear-response
correlation integral. \\
}
\end{figure}

The 0532 Model has only three phase-space dimensions rather than four so that the linear
response simulation is about three orders of magnitude, one thousand times, faster.  The
agreement between the linear-response and directly measured current is likewise excellent,
as is shown in {\bf Figure 5}.  Evidently the ergodic thermostats reproduce both Gibbs'
canonical distribution and linear nonequilibrium perturbations as described by Green-Kubo
theory.

\begin{figure}
\includegraphics[width=4.5in,angle=90.]{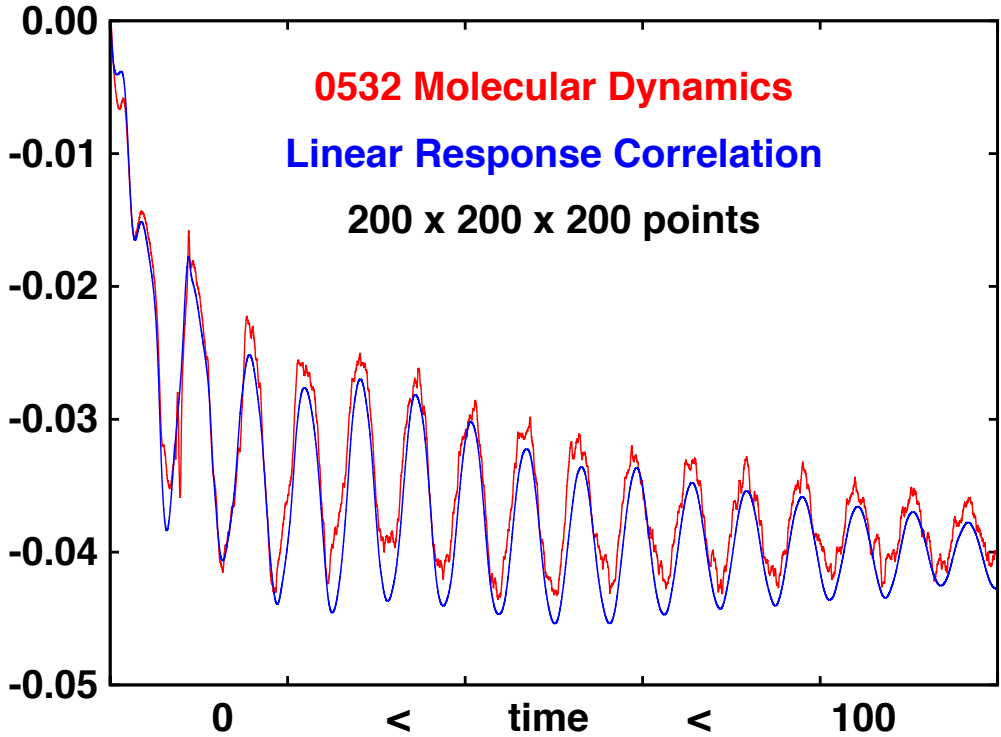}
\caption{
Comparison of the linear-response correlation function with the measured current for the
0532 oscillator at a field strength $\epsilon = 0.10$ .  We show results for
$T = 1 + 0.10 \tanh(q)$ which closely resemble those for
$T = [ \ 1 - 0.10 \tanh(q) \ ]^{-1}$ confirming that $\epsilon = 0.10$ is close to the
linear regime. The three-dimensional Gaussian phase-space integration uses
$200^3$ equally-probable points as the initial states for both the average current and the
correlation integral. \\
}
\end{figure}

\section{SUMMARY and Historical Perspective}

A wide variety of time-reversible thermostats all generate Gibbs' canonical ensemble
through deterministic chaos. When the kinetic temperature varies with coordinate,
the resulting heat current $(p^3/2)$  leads to dissipation, heat transfer, and entropy
change.  The steady loss of comoving phase volume obeys Gibbs' thermodynamic
relations in the extended phase space :
$$
\langle \ (\dot S/k) = (\dot Q/kT) = (\dot \otimes/\otimes) \ \rangle \ ,
$$
where the comoving phase volume includes extensions in the thermostat directions.
These time-averaged relations hold even for the nonergodic Nos\'e-Hoover oscillator :
$$
\langle \  (\dot \otimes/\otimes) \ \rangle = - \langle \  \zeta  \ \rangle
= - \langle \  \zeta (p^2/T) \ \rangle = \langle \ (\dot S/k) \ \rangle \ [ {\rm NH} \ ] \ .
$$

Because the ergodic thermostats all generate Gibbs' canonical distribution they
also  give linear-response relations linking the nonequilibrium currents and thermal
gradients.  We believe that these observations are fundamental to a systematic
exploration of nonequilibrium statistical mechanics through thermostated dynamics.

Our presentday understanding of nonequilibrium systems has its basis in the work
of Boltzmann, the Ehrenfests, Gibbs, and Maxwell. 50 years of numerical work have
provided alternatives to their classic Hamiltonian and stochastic models. Deterministic
reproducibility with dissipative time-reversibility have provided explicit links
between microscopic nonequilibrium molecular dynamics and macroscopic thermodynamics.

Shockwave studies which generate localized far-from-equilibrium states would seem
to be an ideal problem for consolidating these gains in understanding.  Shock
dynamics {\it is} purely Hamiltonian inside the wave and with equilibrium cold and
hot boundaries.  The relaxation times correspond to vibrational collision times. The
nonlinear dependence of transport coefficients and the irreversible nature of the
timelag between forces and fluxes can be measured directly in shockwaves\cite{b18}.
There is a comprehensive listing of nearly all the existing approaches to
nonequilibrium systems in Jepps and Rondoni's review\cite{b19}.  Tools for the
exploration of these problems are close at hand. The only thing lacking in the
shockwave problems is a simple model example like the Galton Board\cite{b11} and
the conducting oscillator studied here.

\pagebreak

\section{Acknowledgment}

We thank John Ramshaw ( Lawrence Livermore Laboratory ) for a continuing series
of thought-provoking emails and for sharing his recent work describing the canonical
thermostating of dynamical systems in very general terms\cite{b20}.  We also
appreciate the Editor's suggestion that the statistical mechanical notation be
outlined ( Section I ) for the mathematical readership of this Journal.

\end{document}